# A PLASMA MODEL FOR RF BREAKDOWN IN ACCELERATOR STRUCTURES*

P. B. Wilson, Stanford Linear Accelerator Center, Stanford University, Stanford, CA 94309

*ABSTRACT*

A plasma model is presented for the formation of "cathode spots" and subsequent crater development near field emission sites on a copper surface in the presence of a strong dc electric field. Adding to previously published models, we propose that the two-stream plasma instability relates the plasma density to its dimensions. Arguments are presented that suggest that the formation and dynamics of such plasma spots are essentially the same phenomenon for both dc and rf fields. A consequence for accelerating structures is that, when such plasma spots are present, they erode the copper surface at a rate on the order of 300 micrograms per Coulomb of ion bombardment current. This erosion rate can have a measurable effect on the rf properties of an accelerating structure after processing at high gradients for a few hundred hours.

## 1 INTRODUCTION

There is an extensive literature on the role played by plasmas in the formation of dc vacuum arcs. Reference[1] provides an excellent summary of the experimental and theoretical status of this field as of 1995. In particular, Chapter 3 on cathode spots will be a cornerstone of this paper. These plasma manifestations consist of "an ensemble of small, luminous spots that move over the surface but instead of resting at the location where the surface is already hot, the spots displace themselves to cold surface areas that must be heated again ([1], p.73). These plasma spots can be imaged by a laser absorption technique (see ([1], Ch. 3, Figs. 22 and 24). Single spots are usually roughly hemispherical in shape (sometimes mushroom shaped), and have diameters mostly in the range $5-20\mu$m. Larger spots seem to be overlapping clusters of smaller spots. During the lifetime of a spot, it excavates a crater by surface heating and melting due to an intense bombardment of plasma ions accelerated by the plasma sheath that separate the body of the plasma from the metallic surface. In addition to crater excavation, this ion bombardment current also vaporizes the cathode material (assumed to be copper) into the plasma. Some of this material emerges from the plasma in the form of a jet of plasma and neutral copper atoms. Microscopic clumps of matter are also ejected from the molten crater by the tremendous pressure (hundreds of atmospheres) produced by the ion bombardment. The major premise of this paper is that plasma spots are also created at a metal surface by intense rf fields, and that these spots share many of the features of the cathode spots observed in dc vacuum arcs. In this paper we concentrate on the physics of these spots on a micron scale length.

*2.1 Ion current Density at the Cathode*

Deferring for a moment a discussion of plasma initiation and the dynamics of its development, we assume the existence of a roughly hemispherical blob of plasma on the copper surface. When a plasma is in contact with a metallic surface, a plasma sheath forms with a potential drop $V_s$ which retards the motion of the more mobile species (electrons) so as to equalize the electron and ion current flow to the wall and thereby keep the plasma charge-neutral [2]. The sheath acts as a space-charge-limited diode for the ion current flowing to the cathode, with a gap spacing which is on the order of the Debye shielding length. Putting in values for the constants in the Child-Langmuir equation assuming copper ions with an average charge 1.85e (from [1], p. 235), the ion current becomes $J_{ion} = 9.3 \times 10^{-9} V_s^{3/2}/\lambda_D^2$. The Debye length is in turn given by $\lambda_D = 7.4 \times 10^3 (KT_e/e)^{1/2}/n^{1/2}$, where $T_e$ is the plasma electron temperature and $n$ the electron density. The potential drop across the sheath is related to the electron temperature by $eV_s/KT_e = \ell n[(J_i/J_e)/(2\pi\alpha)^{1/2}]$, where $\alpha$ is the electron to ion mass ratio ($8.6 \times 10^{-6}$ for copper), and $J_i$ and $J_e$ the counter-flowing ion and electron currents in the sheath region. From [1] values for $V_s$ and $KT_e$ are in the range $kT_e \approx 4-6eV$ and $V_s \approx 18-20V$ for spot currents $20-100A$. Because of the logarithmic dependence in the preceeding equation, the value of $J_i/J_e$ (which we'll need later) is not closely constrained. For $V_s = 19V$ and $kT_e = 5eV$, $J_i/J_e = 0.3$. Using these values of $V_s$ and $KT_e$ in the preceeding expressions for $J_{ion}$ and $\lambda_D$ we obtain

$$J_{ion} \approx 3 \times 10^{-15} n. \qquad (1)$$

This expression relates the ion bombardment current density at the copper surface to the plasma density (unless

*Work Supported by Department of Energy Contract DE-AC03-76SF-00515

otherwise noted, all units are MKS).

## 2.2 Two-Stream Plasma Instability

An interesting experimental observation is that the total current for single spots tends to be nearly constant, independent of spot diameter. This constancy of $I = \pi J r_s^2$ also falls out of a complex theoretical calculation for spot currents of less than about 70A (see [1] Ch. 3, Fig. 62). There is, however, another explanation for $I \approx$ constant: the two-stream plasma instability for an electron current flowing through a plasma with no average ion motion [3]. The instability is convective, meaning that a beam perturbation near the cathode will be amplified as it is carried along by the electron stream, similar to the amplification process in a traveling-wave tube. A good calculation of the convective gain per unit length is currently work in progress. However, a rough estimate of the instability limit can be obtained by solving the dispersion relation in [3] for wave number as a function of frequency to obtain $\tilde{k} = (\omega/v_0)\{1 + j[\alpha - (\omega/\omega_p)^2]^{-1/2}\}$, assuming $\omega/\omega_p < \alpha^{1/2}$. Here $\tilde{k}$ is the complex wave number of a perturbation of real frequency $\omega$, $v_0$ is the electron flow velocity, and $\omega_p$ is the plasma frequency. For the instability to develop, the system length, $L$, must be larger than the longest wavelength of the disturbance by some factor (say 10). Thus, $k_{real} = 2\pi/\lambda, L > 10\lambda$ and therefore $L \geq 20\pi v_0/\omega$. The maximum imaginary part of $\tilde{k}$ (and hence maximum gain) will occur for $\omega/\omega_p \approx \alpha^{1/2}$. Thus, $\omega_p L \geq 20\pi v_0/\alpha^{1/2}$ for instability. The stream velocity can be obtained from $J_e = env_0$. Using $J_e \approx 3J_i$ from the previous section and $J_i \approx 3 \times 10^{-15} n$ from Eq. (1), we obtain $v_0 \approx 6 \times 10^4 m/s$. Using $\alpha = 8.6 \times 10^{-6}$ and $\omega_p = 56\, n^{1/2}$, we obtain $n^{1/2} L \geq 2.5 \times 10^7$. Combining this with Eq. (1), $J_i \approx 1.5/L^2$ at the instability limit, and the current is (assuming $L \approx 2r_s$) $I_i = \pi r_s^2 J_i \approx 20A$, or $I_e \approx 60A$. This is in reasonable agreement with observed single-spot current values, especially given the order-of-magnitude approximations made here.

## 2.3 Crater Formation

To trigger the formation of a plasma, two ingredients are necessary: a source of electrons and a source of gas (See Knobloch[4]). The electrons usually come from field emission from sharp surface irregularities on the scale of one micron, possibly also with further enhancement at areas of absorbed dielectric material. A macroscopic surface field on the order of 20 $MV/m$ with a field enhancement factor of a few hundred has been observed to trigger strong field emission in a dc gap.[1] The necessary gas can come from release of absorbed gas near the field emission site, or possibly from the vapor generated by explosive heating of a microprotusion. We assume here that a plasma then forms in a nanosecond or so with dimensions on the scale of the microprotusion (about half a micron; see Knobloch simulations in [4]).

An important parameter in characterizing thermal effects at metal surface is the heat diffusion distance as a function of time: $x_D = 2(Dt)^{1/2}$, where $D$ is the diffusivity ($D = 1.1 \times 10^{-4} m^2/s$ for copper). In one nanosecond, the heat generated at the surface will penetrate 0.3 microns into the metal. If we apply a source of heat from plasma ion bombardment using $J_i \approx 0.4/r_s^2$ from Sec. 2.2 and taking $V_s = 20V$, $P_A = V_s J_i \approx 8/r_s^2$. For a spot with $r_s = 0.5\mu m$, and assuming that half the power goes into bringing the copper surface to the melting point, then $P_A \approx 1.5 \times 10^{13} W/m^2$. In the semi-infinite solid limit (valid for $x_D < r_s$), the time required to bring the surface to the melting point ($\Delta T = 1060°C$) is $t_{melt} = (\pi/D)(K\Delta T/2P_A)^2$ where $K$ is the thermal conductivity ($380 W/m$-$K$ for copper). Putting in numbers, $t_{melt} \approx 0.01\, ns$. This is small compared to the plasma formation time ($\sim 1 ns$). The conclusion is that the copper surface beneath the plasma will already be melting by the time the plasma is fully formed.

After formation of the initial molten site, assuming a hemispherical crater, further melting proceeds following

$$2\pi r^2 \rho (L_h + C_h \Delta T) dr \equiv Ar^2 dr = P_0 dt$$

where $\rho = 9 g/cm^3$ is the copper density, $L_h = 212 J/g$ is the latent heat of melting and $C_h = 0.42 J/g$-$K$ is the specific heat. For an $I_e = 50A$ per spot ($I_i \approx 15A$), the available input power is $P = r_s I_i \approx (20V)(15A) = 300W$. We will again assume about one-half of this is available for crater melting. The above equation is readily integrated to obtain the crater growth from its initial size of about a micron, $r - r_0 = (3P_o t/A)^{1/3}$, where $A = 3.7 \times 10^4 J/cm^3$. For $t = 1, 10$ and $50 ns, r \approx 3, 5$ and 10 microns for small $r_o$. During the melting process, the ion bombardment pressure ($\sim 0.1 GPa$) pushes material from the crater to form a rim, where it cools and solidifies. At some time between 10 and 50 ns, called the spot residence time, the whole process becomes unstable and one or more new smaller plasmas spots tend to form at sharp irregularities on or near the crater rim. Sometimes the old spot may simply extinguish. There may be an explosive ejection of any remaining molten material due to exposure to the strong surface fields.

## 2.4 Material Erosion Rate

Some material is ejected from the crater region by explosive events, but the major effect acting to remove material is evaporation of copper into the $T \approx 3000°K$ plasma. The incident ion energy is on the order of 20 $eV$, and only a fraction of this (a few $eV$) is needed to vaporize a copper atom. If each incident ion with a charge of 1.85$e$ were to evaporate two copper atoms, the net erosion rate would be $(1.85 \times 1.6 \times 10^{-19})^{-1} = 3.4 \times 10^{18}$ particles per Coulomb divided, by $(6 \times 10^{23}/63.5) = 9.4 \times 10^{21}$ particles per gram to give a net erosion rate of 360 $\mu g/C$ of ion bombardment current, or about 100$\mu g/C$ of electron current. Measured erosion rates vary from $20-500\mu g/C$, but seem to average around $100\mu g/C$ [1].

*2.5 Relation Of Erosion Rate To Structure Phase Error*

If copper is being eroded at the disk types of an accelerating structure by plasma etching, at some point sufficient material will have been removed to have a measurable effect on the accelerating structure rf characteristics. For example, the phase shift per cell in a $2\pi/3$ mode structure is $\Delta\phi = (\lambda/3)dk = (\lambda/3)d\omega/v_g$, or $\Delta\varphi = (2\pi/3)(d\omega/\omega)/(v_g/c)$. From perturbation theory, $d\omega/\omega = \epsilon_o E_s^2 \Delta V/4U$, where $E_s$ is the surface field, $\Delta V$ is the volume removed and $U$ the stored energy per cell, $U = (\lambda/3)u = (\lambda/3)E_a^2/s = (\lambda/3)E_s^2/f^2s$. Here $f = E_s/E_a$ is the ratio of surface field to average accelerating gradient, and $s$ is the elastance per meter. Combining these two expressions gives $d\omega/\omega = 3\epsilon_o f^2 s \Delta V/4\lambda$. Substituting in the expression for $\Delta\varphi$,

$$\Delta\varphi = \frac{(\pi/2)\epsilon_o s f^2}{\lambda(v_g/c)} \Delta V. \quad (2)$$

Substituting in $s = 900 \times 10^{12} V/C$-$m$, $v_g/c = 0.10$, $f = 2.5$ and $\lambda = .026m$, we obtain $\Delta\varphi \approx 3 \times 10^7 (\Delta V/m^3)$.

Assume that about half of the eroded material returns near the crater site and that the other half tends to land well away, with an equal chance of ending up in an electric or magnetic surface field region. (In an rf field, any charged material ejected into the vacuum at low velocity has an equal chance of moving toward or away from the source). For a net erosion rate of $50 \times 10^{-6} g/C$ and a spot electron current of 40 A, material is removed at a rate of about $2 \times 10^{-10} m^3/s$. This produces a phase shift of $6 \times 10^{-3} t_s$, where $t_s$ is the total spot residence time.

Near the front end of an accelerating structure tested is the ASTA facility at SLAC, a phase shift of $6 \times 10^{-3}$ radians per cell due to iris tip erosion was measured after processing at high power for about 400 hours. An integrated spot residence time of about one second would produce the measure phase shift. At a repetition rate of 60 Hz and a pulse length of $150ns$ the accumulated pulse time is 13 seconds. Thus, a plasma spot needs to be present on about 8% of the pulses to explain the erosion. Also, there should be about $(1s)/(150ns \times 60Hz) = 1 \times 10^5$ craters per iris. The total number of full rf breakdown events (with measurable reflected power) was only a few hundred per iris with heavy damage during the 400 hour processing period in ASTA. Thus, on pulses when cathode spots were present, an rf breakdown event was triggered less than 1% of the time according to the plasma spot model.

## 3 SUMMARY

We have proposed that the formation and behavior of plasma spots on a copper surface at high field strengths is essentially the same phenomenon for both dc and rf fields. For the rf case, Knoblock's simulation [4] shows that a dense plasma, with dimensions on the order of a micron, readily forms near a field emission site on a very short time scale (a few rf cycles), if a source of gas is also available at the site. The vaporization of a microprojection could in itself be a gas source.

Once the plasma spot has formed, the copper surface the plasma is shielded from the rf field, and the dynamics of crater growth should proceed in a similar manner for both dc and rf fields, at least until the (possibly) explosive end-life of the crater, when the plasma extinguishes or jumps to a new location on or near the crater rim.

During the spot lifetime, a simple calculation shows that a hemisphere of plasma is not significantly perturbed by an rf field. Since the plasma inside the hemisphere is neutral, rf field lines must terminal on a surface charge layer with density $\epsilon_o E_s$ per unit area. The rf field will alternately peel off, at most, an electron or ion charge $2\pi r^2 \epsilon_o E_s$ per rf cycle, giving a current $I = dq/dt = \epsilon_o E_s r^2 \omega$. At 11 GHz, with $E_s = 2 \times 10^8 V/m$ and $r_s = 10\mu m$, this current is $\sim 0.01A$. This is negligible compared to the electron current flow of $10-50A$ for typical (single) plasma spots.

The conditions for spot end-life, and for spot multiplication and location jumping, may involve rf considerations. Occasionally these effects avalanche in such a way as to inject a large amount of charge into the vacuum in a very short time, which rapidly dumps the energy stored in an accelerating cell or cavity to produce a full-blown breakdown event with reflected power. Plasma spots, however, are seen as existing independently of such breakdown events, and are in fact required to be present on most rf pulses, when processing close to breakdown, to explain observed erosion rates and crater counts. It remains to produce a satisfactory model which connects the plasma spots discussed in this paper with the observed features of full rf breakdown in structures and cavities, and to explain the scaling of the breakdown gradient with frequency and pulse length.